# Charge Transfer States in Donor-Acceptor Bulk-Heterojunctions as Triplet-Triplet Annihilation Sensitizer for Solid-State Photon Upconversion

Maciej Klein,[1,2*] Alexander R. Ireland,[3] Evan G. Moore,[3*]

Dennis Delic[4] and Ajay K. Pandey[1,2*]

[1] School of Electrical Engineering and Robotics, Queensland University of Technology, Brisbane, QLD 4001, Australia

[2] Centre for Materials Science, Queensland University of Technology, Brisbane, QLD 4001, Australia

[3] School of Chemistry and Molecular Biosciences, The University of Queensland, Brisbane, QLD 4072, Australia

[4] Defence Science and Technology Group, Edinburgh, SA 5111, Australia

*Correspondence to: maciej.klein@qut.edu.au, egmoore@uq.edu.au, a2.pandey@qut.edu.au

Photon interconversion in semiconductors is of fundamental importance for digital imaging and quantum sensing. Nonlinear processes such as triplet-triplet annihilation (TTA) offer photon upconversion (UC) at yields desired for solid-state optoelectronic devices. Here, we present a multilayer molecular system where a near-infrared (NIR) photosensitizer facilitates robust photon UC. A molecular stack of 2,4-Bis[4-(N,N-diisobutylamino)-2,6-dihydroxyphenyl] squaraine (DIB-SQ): [6,6]-Phenyl C61 butyric acid methyl ester (PCBM) is optimized for UC by fine tuning the PCBM loading to engineer charge transfer states that facilitate further triplet generation. This composite NIR photosensitizer layer, at a 1:3 blend ratio in heterojunction with rubrene, drives triplet population density to levels desired for effective TTA. When paired with a fully optimized annihilator layer of tetraphenyldibenzoperiflanthene (DBP) doped rubrene, the NIR sensitizer produces a photon upconversion quantum yield ($\Phi_{UC}$) of 1.36% at 690 nm, with a significantly low excitation intensity threshold for the onset of the linear regime $I_{th} = 60.5 \ mW/cm^2$. Time-resolved photoluminescence, transient absorption spectroscopy, and magnetic field dependent photoluminescence measurements reveal a detailed balance of photoexcited states and formation of charge transfer states of triplet character ($^3$CT), which work in tandem with molecular states to sensitize the triplet state ($T_1$) of rubrene. This approach of harnessing near-infrared photons presents promising avenue for advancing solid-state photon interconversion.

# 1. Introduction

Advanced imaging and communication technologies such as Lidar, night vision, and quantum encrypted communication networks often utilize the near-infrared (NIR) region of the electromagnetic spectrum[1-4]. The direct harvesting of low-energy photons at room temperature in NIR absorbing low-band-gap semiconductors is a topic of intense research, given their technological importance to Internet of Things (IoT), automotive, and automation industries[5-7]. An alternative approach is to convert NIR photons into the visible band using exotic nonlinear processes such as Triplet-Triplet Annihilation (or triplet fusion, TTA), commonly referred to as photon upconversion (UC). In contrast to other anti-Stokes processes for UC, the unique spin-dependent phenomenon of TTA is accessible under incoherent and low-intensity illumination conditions[8], making it particularly well-suited for solid-state photon interconversion and photodetector technologies[9-11].

While most UC systems have been realized in solution phase, solid-state demonstrations play a pivotal role in achieving technological significance, particularly for optoelectronic devices[11]. Despite the maximum theoretical limit for TTA-based upconversion is 50%[12], solid-state systems typically exhibit low quantum yields ($\Phi_{UC}$ <0.3%)[7, 13-14] primarily due to nonradiative losses from sensitizer aggregation and energy losses during intersystem crossing (ISC). Direct triplet-triplet energy transfer (TET) between the sensitizer and annihilator has allowed to mitigate the detrimental effect of ISC[15-16], spurring interest in inorganic sensitizers such as lead halide perovskites[17] and 2D monolayer Transition Metal Dichalcogenides (TMDs) nanostructures[14, 18], leading to $\Phi_{UC}$ of ~1.1% in an ultrathin bilayer $MoSe_2$/rubrene system[14]. The advantage of TTA as a stepping stone for UC is further realized by co-existence of charge transfer (CT) states-mediated triplet energy transfer in type II organic heterojunctions. A model system based on rubrene/$C_{60}$ serves as an excellent example of this synergy, where energy-upconverted electroluminescence was first demonstrated in a solid-state device operating at room temperature[19-21].

More recently, triplet energy transfer mediated by charge recombination at organic heterojunctions comprising non-fullerene acceptors (NFAs) and rubrene has attracted significant attention in the context of photon upconversion. Further exploitation of CT states formed at the NFA sensitizer/annihilator interface in organic bilayer heterojunctions has led to recent increase in NIR-to-visible photon upconversion of up to ~2.5%[22-23]. This strategy requires increased interfacial contact between sensitizer and annihilator, and subsequently bulk-heterojunction (BHJ) architecture for UC was proposed[24]. The BHJ concept has been further extended from binary to ternary system by including the sensitizer layer, enabling UC through charge separation at the donor-acceptor interface of sensitizer and subsequent formation of charge transfer states between the energy donor and annihilator[25]. However, the prospect of treating molecular triplet states as distinct intermediate states that can further enhance the process of triplet generation for efficient photon UC is not fully realised yet. Within this context, the scope of fullerene compounds in fine tuning the $^3$CT states formation in desired BHJ sensitizers has remained unexplored for NIR to visible photon UC.

On the other hand, TTA has been proposed as a beneficial design principle to reduce turn-on voltages, enhance emission brightness, and improve the external quantum efficiency of electroluminescence in organic light-emitting diodes (OLEDs) and light-emitting field-effect transistors (LEFETs)[20, 26]. As an early proposers of this concept, our group has successfully extended the triplet exciton harvesting frameworks in planar heterojunctions[21] as well as in ternary bulk-heterojunction (BHJ) systems[27] for applications in NIR photodetectors and electroluminescence devices[28].

Through this contribution, we further expand the framework of triplet harvesting using TTA in conjunction with charge transfer states of triplet character ($^3$CT) in carefully engineered energy cascades of organic electron-donors (D) and fullerene-based electron-acceptors (A). The advantage of our approach combines multistep functions of $^3$CT formation and energy transfer in a quaternary like BHJ system, where a solution processed binary BHJ of DIB-SQ:PCBM is used as a triplet sensitizer to amplify TTA in DBP doped Rubrene BHJ- prepared by vacuum co-deposition (detailed in the experimental section).

Here, it is important to highlight that our approach specifically relies on exploiting the formation of $^3$CT in fullerene-based acceptors. Such formation has been classically treated as a terminal loss channel for photogeneration in low-bandgap donor-acceptor systems[5, 29]. In contrast, our work considers the $^3$CT states formation, realized in the model fullerene based BHJ system of DIB-SQ:PCBM, as an important intermediate step to increase triplet excitons population in annihilator. The energetically favorable design of this cascade allows triplet excitons to subsequently undergo TTA, enabling onset of the linear regime for NIR-to-visible photon UC at relatively low excitation thresholds.

Firstly, we optimized the formation of triplet excitons desired for TTA induced UC by fine tuning the binary BHJ composition of DIB-SQ:PCBM as a suitable sensitizer. The evolution of $^3$CT population in DIB-SQ:PCBM as a function of the loading of PCBM was therefore investigated (from 1:0 to 1:5). The 1:3 blend ratio was found optimal for maximizing the population of rubrene triplet states $T_1$ through sensitization by $^3$CT states. Optical spectra of the DIB-SQ:PCBM reveal a blue shift at higher loading of fullerene, a clear indication of reduced aggregation in DIB-SQ. This strategy provides fine tuning of the formation of charge transfer states with respect to TTA annihilator and photoemitter layer of rubrene, ultimately enabling TTA-mediated solid-state photon UC (**Fig. 1a**). The underlying photophysics propelling the UC mechanism is fully characterized by time-resolved and steady-state spectroscopic measurements, which identify molecular triplets as distinct intermediates that facilitate triplet energy transfer across the sensitizer/annihilator interface. Further evidence of TTA-defined upconversion is provided by the magnetic field effect on the photoluminescence.

## 2. Results and Discussion

As stated earlier, the upconversion sensitizer stack proposed in this study consists of a donor–acceptor bulk-heterojunction integrated into a planar junction with an annihilator layer (**Fig. 1a**). A DIB-SQ squaraine molecule (2,4-Bis[4-(N,N-diisobutylamino)-2,6-dihydroxyphenyl] squaraine) acting as the electron-donor and a PCBM ([6,6]-Phenyl C61 butyric acid methyl ester) acting as the electron-acceptor forms the NIR sensitive BHJ layer. Either a pristine rubrene molecule, or rubrene doped with DBP[30] is selected as the annihilator cum photoemitter layer (**Fig. 1b,c**). The sensitizer layer was spun-cast onto a glass substrate, followed by physical vapor deposition under vacuum of the annihilator layer, forming a photon upconversion system, hereafter denoted as DIB-SQ:PCBM/rubrene and DIB-SQ:PCBM/rubrene:DBP. The energy levels of all materials used in this study are shown in **Fig. 1b**.

The choice of DIB-SQ stems from the characterized high absorption coefficients and narrow absorption bands in the NIR region of squaraine derivatives [31]. In the solid-state, these molecules typically form either J- or H-type aggregates, depending on their structure and film preparation conditions[32]. Squaraines have been extensively studied in donor-acceptor bulk-heterojunction solar cells, particularly in blends with fullerene derivatives (typically up to 1:6 squaraine:fullerene weight ratio)[33-34]. Moderate power conversion efficiencies observed in these systems have been attributed to the $^3$CT states formation as a prevalent route for non-radiative recombination and a terminal loss channel[35-38]. We deliberately selected DIB-SQ:PCBM system due to the prevalence of relatively high non-geminate recombination due to $^3$CT formation, and to trigger a triplet energy transfer from $^3$CT to $T_1$ of rubrene.

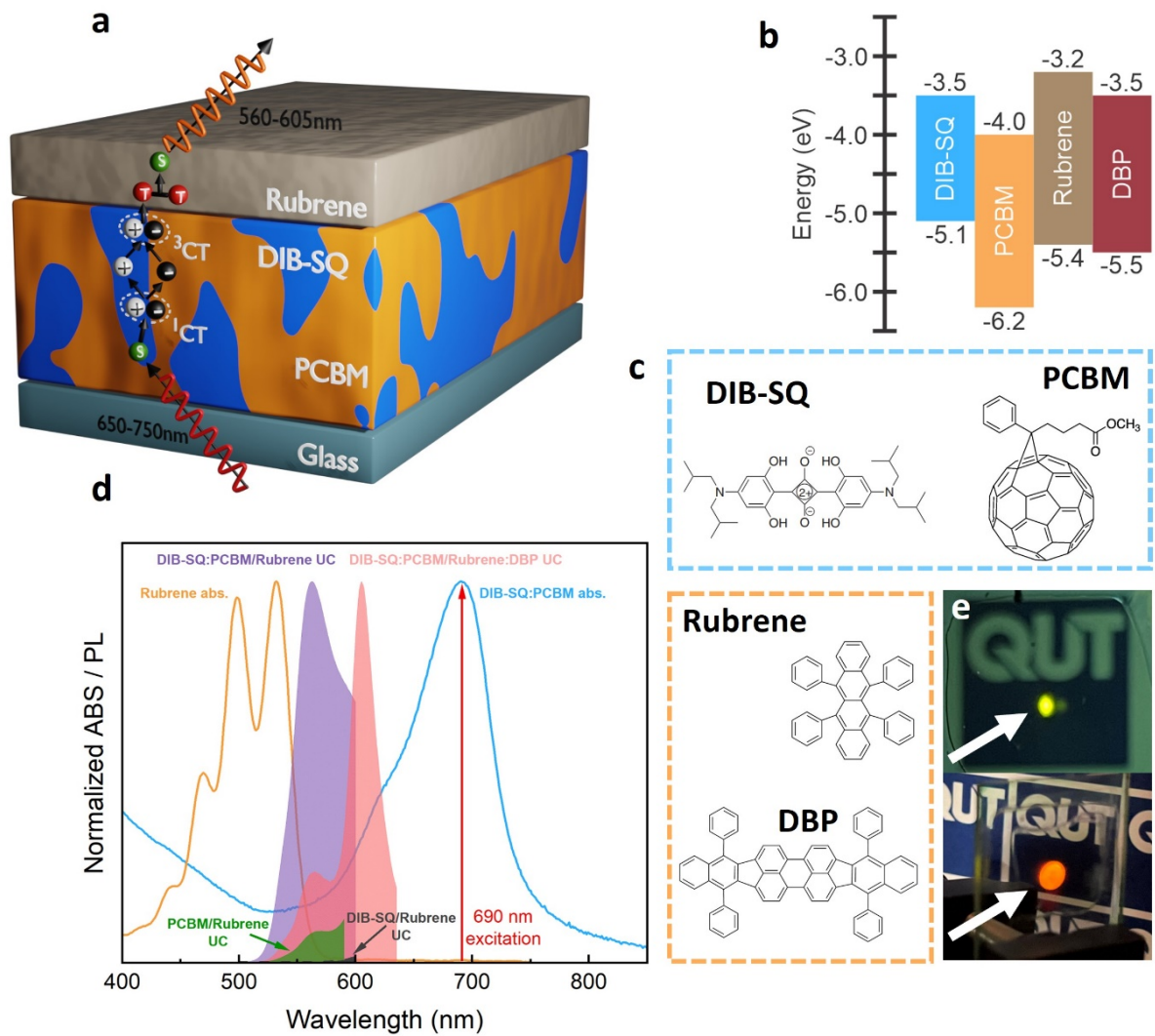

**Figure 1. Scheme of the photon upconversion system and spectroscopic characterization.** (a) Schematic of the UC system showing the separation of charges at donor-acceptor interface (DIB-SQ and PCBM, respectively) followed by triplet charge transfer state formation ($^3$CT) and subsequent triplet energy transfer to the annihilator (rubrene), where S – singlet, T – triplet excitons, CT – charge transfer state. (b) Energy level diagram of DIB-SQ[39], PCBM[40], rubrene[22], and DBP[30], with energy values obtained from the literature. (c) Chemical structure of the donor-acceptor sensitizer system (DIB-SQ and PCBM, respectively) and annihilator (rubrene) doped with DBP. (d) Absorption (abs.) spectra of rubrene and DIB-SQ:PCBM (1:3) (orange and light blue solid lines, respectively). Upconverted (UC) photoluminescence emission from a system with i) neat DIB-SQ sensitizer and rubrene annihilator (DIB-SQ/rubrene UC, gray shaded area), ii) neat PCBM sensitizer and rubrene annihilator (PCBM/rubrene UC, green shaded area), iii) bulk-heterojunction sensitizing layer DIB-SQ:PCBM(1:3)/rubrene UC (purple shaded area), and iv) the same sensitizing layer with rubrene:DBP annihilator (DIB-SQ:PCBM(1:3)/rubrene:DBP UC, red shaded area), all under 690 nm laser excitation with 710 mW/cm$^2$ intensity. Note, DIB-SQ/rubrene and PCBM/rubrene UC are normalized with respect to the maximum of DIB-SQ:PCBM/rubrene UC signal. (e) Photographs of the UC showing yellowish green emission (top) from the DIB-

SQ:PCBM(1:3)/rubrene, and orange emission from DBP doped rubrene DIB-SQ:PCBM(1:3)/rubrene:DBP, under 690 nm laser excitation. No filter was used for rubrene:DBP and a 600 nm low pass filter was used for rubrene only device. The white arrows are guide to the eye to indicate the emission spot.

In solution phase, DIB-SQ molecules exhibit a typical narrow absorption band, with a maximum at 655 nm and a full width at half maximum (FWHM) of 28 nm, as recorded in chlorobenzene (**Fig. S1a** in Supplementary Information). In contrast, the thin film absorption spectrum shows a characteristic broad peak with two distinctive features: one red-shifted and the other blue-shifted relative to the monomer absorption peak, with maxima around 625 nm and 725 nm, $FWHM = 200\ nm$, attributed to the charge transfer H-aggregate[41]. Upon the addition of PCBM, this aggregate coupling is reduced, and the peak around 688 nm ($FWHM = 82\ nm$), originating from monomer squaraines, becomes dominant. As expected, a strong quenching of the squaraine photoluminescence (PL) around 750 nm was observed (**Fig. S1b** in Supplementary Information), due to formation of CT states and efficient electron transfer to fullerene. Such quenching in blends with high PCBM content can partially be attributed to the phase-separation-induced squaraine aggregation (self-quenching)[42].

Photon upconversion experiments include excitation of the DIB-SQ:PCBM sensitizer layer with an incoming NIR light (690 nm). The photon emission of rubrene in the 550-650 nm range is observed, exhibiting a distinct apparent anti-Stokes photoluminescence shift characteristic of upconversion processes (**Fig. 1d**). The UC photoluminescence overlaps well with rubrene or rubrene:DBP photoluminescence when directly excited at 450 nm (**Fig. S2** in Supplementary Information ) and is visible to the naked eye (**Fig. 1e**). Conversely, devices employing either neat DIB-SQ or PCBM sensitizer layers exhibited significantly reduced upconversion performance. No UC emission was observed from DIB-SQ/rubrene devices, while PCBM/rubrene ones showed weak emission only under coherent excitation at laser intensities above 100 mW/cm². These results suggest that the performance of devices utilizing neat DIB-SQ or PCBM layers is severely limited, likely due to inefficient ISC-driven triplet formation in these films[43]. In contrast, the DIB-SQ:PCBM blend enabled UC emission even under low-intensity incoherent excitation, underscoring the pivotal role of CT states in the energy transfer process, facilitating effective population of triplet states in rubrene. Optimization of the DIB-SQ:PCBM blend composition (from 1:1 to 1:5 weight ratio) revealed the optimal combination for overall light output, defined as the ratio of upconverted photoluminescence intensity (excited at maximum sensitizer absorption) to direct rubrene photoluminescence (excited at 465 nm), to be 1:3, denoted as SQ:PCBM (1:3) – **Fig. S3** in Supplementary Information. Consequently, the following sections focus mainly on the 1:3 BHJ composition.

The optical properties of the proposed UC system were examined using excitation wavelength-dependent upconverted photoluminescence measurements. The action plot of the integrated UC emission aligns closely with the absorption spectrum of the DIB-SQ:PCBM (1:3)/rubrene system, often referred to as the symbatic response[44-45]. This overlap confirms that UC originates from light absorbed by the sensitizer layer, followed by excited state recombination in the bulk[46] (**Fig. 2a**). The upconversion kinetic studies revealed the generation of rubrene singlet states on a timescale of $\tau_g = 1.4\ \mu s$, with a lifetime of about $\tau_d = 14.7\ \mu s$ (**Fig. 2b**). The data is well described by mono-exponential growth and decay functions: $I_{UC}(t) \propto e^{-t/\tau_d} - e^{-t/\tau_g}$, where $I_{UC}$ is the upconversion photoluminescence intensity.

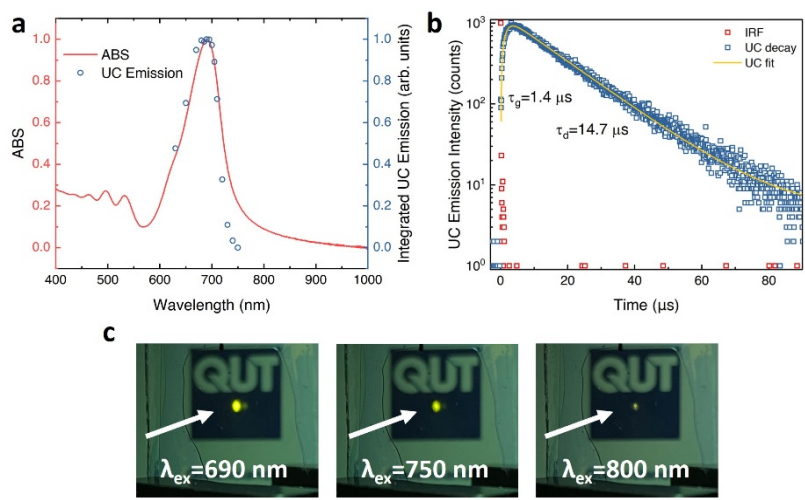

**Figure 2. Upconversion in a rubrene-only based system.** (a) Action plot of the excitation wavelength-dependent energy upconverted photoemission and corresponding absorption spectrum of the DIB-SQ:PCBM (1:3)/rubrene UC system. (b) A semi-log plot of the TRPL kinetics for rubrene emission under 640 nm pulsed laser excitation - $\tau_g$ and $\tau_d$ denote UC growth and decay time, respectively. (c) Photographs of the UC emission under 690, 750, and 800 nm laser excitation (taken through a 600 nm short-pass filter).

Photographs of the rather weak UC emission under 690, 750, and 800 nm laser excitation are shown in **Fig. 2c**. The upconversion quantum yield, $\Phi_{UC}$, of rubrene-based devices is typically low, primarily due to the intrinsically low photoluminescence quantum yield (PLQY) of rubrene in the solid state. This low PLQY (usually below 5%) is attributed to the formation of an amorphous phase, molecular aggregation, and possibly singlet fission (SF)[47-49]. To overcome these limitations, rubrene is often doped with the singlet energy collector tetraphenyldibenzoperiflanthene (DBP). DBP is known to harvest singlet energy from rubrene via Förster Resonance Energy Transfer[48] (FRET), significantly increasing its PLQY.

To enhance the performance of proposed upconversion system, rubrene film was doped with 0.5 vol. % DBP via co-evaporation. The observed UC emission peak shifted from 562 nm in the device with pristine rubrene annihilator to 605 nm in the rubrene:DBP blend (**Fig. 1d** and **Fig. S2** in Supplementary Information). The upconversion quantum yield, defined as the number of UC photons emitted divided by the number of photons absorbed by the sensitizer (with the maximum theoretical efficiency of 50%)[12], was estimated by using a relative method. The rubrene:DBP film, with a PLQY of $47.5 \pm 2.4\%$, determined by using an integrating sphere and the de Mello method[50], was used as a reference standard. The saturated upconversion yield of the DIB-SQ:PCBM (1:3)/rubrene:DBP device, with an irradiation intensity above 69.7 mW/cm$^2$ was $\Phi_{UC} = 1.36\%$ (**Fig. 3a**).

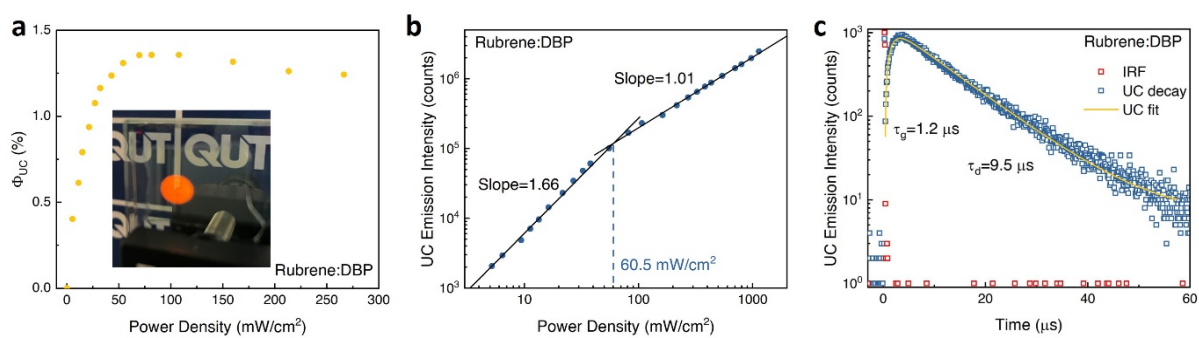

**Figure 3. Upconversion performance in rubrene:DBP based system.** (a) The UC quantum yield of the DIB-SQ:PCBM (1:3)/rubrene:DBP device as a function of excitation power density. (b) A log-log plot showing the dependence of the upconverted photoluminescence intensity on the incident light intensity under 690 nm laser irradiation. (c) A semi-log plot of the TRPL kinetics for rubrene:DBP emission under 640 nm pulsed laser excitation.

The excitation intensity dependence of the UC emission, measured under 690 nm continuous-wave (cw) laser illumination, revealed two distinct regimes, one with a slope of 1.66 at lower excitation intensities and another at higher intensities with a slope of 1.01 (**Fig. 3b**). The excitation light intensity at the transition point between the sub-quadratic and linear dependence regions, (typical for TTA-based UC systems, and called the threshold intensity, $I_{th}$), was determined to be $I_{th} = 60.5 \ mW/cm^2$. This corresponds to an excitation density of approximately $1.28 \times 10^{17} \ s^{-1} \ cm^{-2}$ (at 690 nm). The $I_{th}$ value is among the lowest reported for a solid-state UC systems[13] and comparable to those recently obtained for rubrene/Y6 and rubrene/ITIC-Cl bilayers and BHJs[22, 24], and PYIT1:PBQx-TCl:rubrene BHJ system[25].

Upconversion dynamics measurements revealed the generation of DBP singlet states on a timescale of $\tau_g = 1.2\ \mu s$, followed by a mono-exponential decay with a lifetime of $\tau_d = 9.5\ \mu s$ (**Fig. 3c**).

To gain deeper insight into the charge carrier kinetics underlying the UC process and to reveal the nature of the excited states involved, femtosecond and nanosecond-transient absorption (TA) measurements were carried out, with probe pulses spanning the visible and NIR spectra. A neat DIB-SQ film is characterized by positive signatures below 545 nm and in the NIR (with maxima around 1030 and 1180 nm), which are assigned to the $S_1 \rightarrow S_n$ singlet excited state absorption (ESA)[43, 51] – **Fig. 4a.** A broad negative band spanning from 550 to 800 nm is attributed to photobleaching of the ground state (GSB), with its falling edge around 750 nm coinciding with a negative stimulated emission (SE) feature that closely resembles the steady-state fluorescence spectrum. The small addition of PCBM (DIB-SQ:PCBM (1:0.5)) slightly quenched the fluorescence signal and altered the NIR response, making it broader and more pronounced – **Fig. S4** in Supplementary Information. In the DIB-SQ:PCBM (1:3) blend, the squaraine ESA signal in the visible region is redshifted by about 100 nm and the SE is significantly quenched due to the interaction between donor and acceptor molecules[51] – **Figs. 4b** and **4c**. Additionally, the excited state absorption features in the NIR (around 1030 and 1180 nm) decay with a time constant of 1.2 ps (as extracted from global analysis). Meanwhile, strong polaron features with maxima around 1120 and 1350 nm emerge[51], indicating the formation of CT states, where the electron resides on PCBM and the hole on the squaraine molecule. **Fig. 4d** shows the ESA decay monitored at 1180 nm, followed by a rise of the polaron signature at 1350 nm. Two additional features are visible on the nanosecond timescale - **Fig. S5a** in Supplementary Information. A positive signal at 720 nm is assigned to $T_1 \rightarrow T_n$ triplet absorption in PCBM[52] (denoted as T$_{PCBM}$) and a weak SE feature in the 860-920 nm range, which may be caused by the emission from CT states[53-54]. The slow decay of the polaron feature monitored at 1350 nm is followed by a rise in the T$_{PCBM}$ population at 720 nm, indicating the recombination of CT states via molecular triplets in the acceptor – **Fig. S5b** in Supplementary Information.

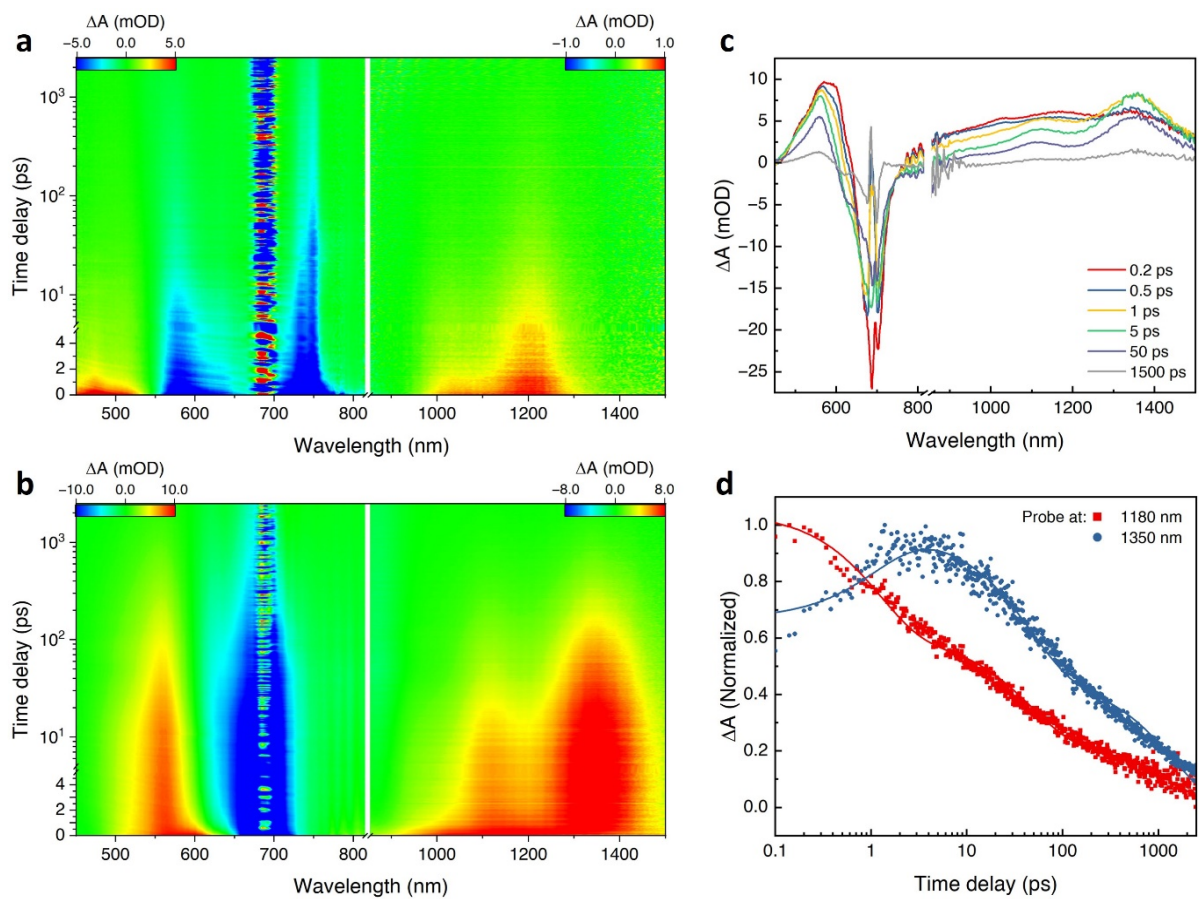

**Figure 4. Transient absorption spectroscopy of the donor-acceptor sensitizer.** 2D femtosecond-transient absorption contour map of (a) DIB-SQ and (b) DIB-SQ:PCBM (1:3) films in the visible to near-infrared region. (c) Femtosecond-TA spectra of DIB-SQ:PCBM (1:3) film at different probe delay times. Some scattered light from the excitation pulse was observed at 688 ± 10 nm. (d) Normalized TA decay kinetics of DIB-SQ:PCBM (1:3) film monitored at 1180 and 1350 nm. The solid lines represent the associated kinetic fitting curves extracted from global analysis.

In D-A systems, the oscillator strength of optical transitions from the ground state to the CT states is generally weak, making them typically undetectable in steady-state absorption spectra[53, 55]. However, these transitions are possible and have been observed in the photocurrent spectra of organic photovoltaic (OPV) devices, suggesting that the direct photoexcitation of CT states can contribute to the generation of free charge carrier[56-57]. Therefore, the presence of CT states in the blend is further evidenced by absorption and high-sensitivity external quantum efficiency (EQE) measurements of OPVs with DIB-SQ (only), PCBM (only) and DIB-SQ:PCBM (1:3) active layers (**Fig. S6a** in Supplementary Information). The CT states energy of $E_{CT} = 1.44\ eV$ was determined by fitting the low-energy absorption tail of the EQE spectrum and the high-energy tail of the emission spectrum extracted from ns-TA (**Fig. S6b** in Supplementary Information and supplementary note)[56].

Upon deposition of rubrene onto the BHJ sensitizer blend to form the DIB-SQ:PCBM (1:3)/rubrene UC platform, nanosecond-TA measurements reveal a complex interplay between the molecular states of the chromophores and charge transfer species, which leads to the upconverted emission (**Fig. 5** and **Fig. S7** in Supplementary Information). The long-lived, distinct signatures with overlapping positive and negative amplitudes between 430 and 570 nm, along with an additional negative feature around 615 nm, can be identified (**Fig. 5a**). The positive peak at 510 nm is characteristic of $T_1 \rightarrow T_n$ triplet excited state absorption in rubrene[58], while the negative peaks at 465, 495, and 530 nm align with the rubrene steady-state absorption spectrum (**Fig. S8** in Supplementary Information). Such ground state bleach signatures can be observed particularly for the long-lived intermediate species. Features at approximately 560 and 615 nm correspond to the photoluminescence of rubrene. Although these features, which represent stimulated emission, should appear negative, the overall positive TA signal at 560 nm suggests a higher oscillator strength for excited state absorption of the encounter complex with a singlet character, $^1$(T-T)$_{Rub}$, formed upon triplet-triplet annihilation in rubrene[20]. During the collision of two triplet states, depending on the relative orientation of the spins, an encounter complex with either singlet, triplet, or quintet character can be formed, but only the singlet spin configuration of the triplet pair leads to delayed fluorescence[59].

The corresponding nanosecond-TA kinetics monitored at 890 and 1350 nm show the depopulation of CT states occurring within the instrument response function of approximately 100 ps. This is followed by a rise (and subsequent slow decay) of the molecular triplet state population in rubrene and PCBM, at 510 and 720 nm, respectively, along with a rubrene bleach at 530 nm (**Fig. 5b**). A fit to the kinetics at 510 nm yields a rubrene triplet population rise of approximately 5.2 ns and a lifetime of about 1.4 μs. Although rubrene singlet states are not directly populated by photoexcitation, they should still be considered in the triplet formation process due to singlet fission, where a singlet exciton splits into two low-energy triplet excitons[60]. In neat rubrene films, the triplet population associated with a fission rises in ~1.8 ps, and the corresponding TTA-induced delayed fluorescence has a lifetime of ~280 ns. In rubrene/C$_{60}$ bilayers, these times are slightly longer, at 8 ps and 590 ns, respectively[20]. Therefore, the contribution of singlet fission to the observed long-lived upconverted emission can be safely neglected.

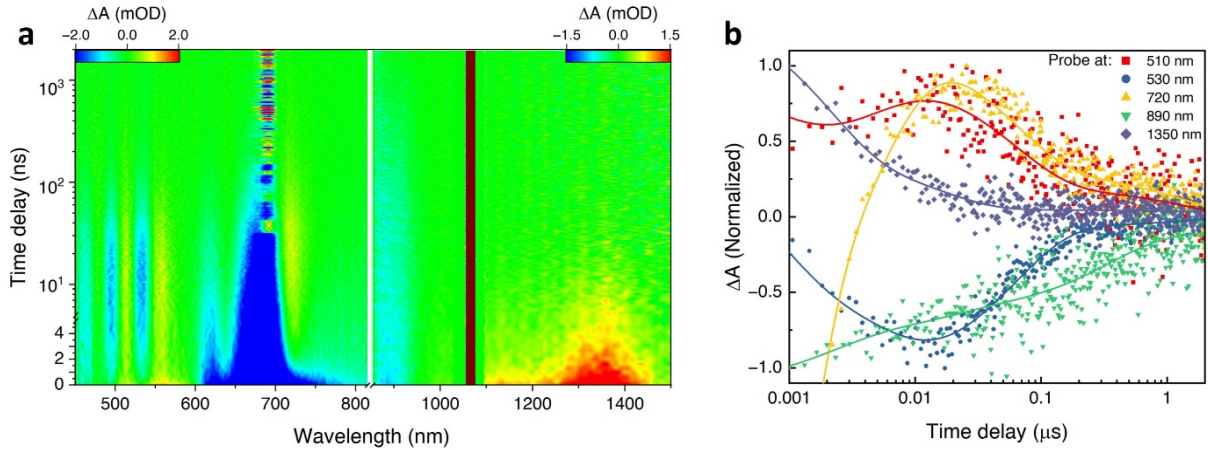

**Figure 5. Energy transfer kinetics.** (a) 2D nanosecond-transient absorption contour map of DIB-SQ:PCBM (1:3)/rubrene system excited at 688 nm. Some scattered light from the excitation pulse was observed at 688 ± 10 nm. (b) TA kinetic traces showing energy transfer to rubrene triplet ($T_{Rub}$ probed at 510 nm) mediated by charge transfer states formed at donor-acceptor interface (CT - probed at 1350 nm, and their emission $CT_{PL}$ – probed at 890 nm) and molecular triplet state of PCBM ($T_{PCBM}$, probed at 720 nm), and rubrene ground state bleaching ($GSB_{Rub}$, probed at 530 nm). The solid lines represent the associated kinetic fitting curves extracted from global analysis.

The manifold of charge transfer (CT), charge separated (CS) states, and molecular triplet states in either donor or acceptor provides a unique energy exchange window that facilitates the population of triplet states in rubrene, followed by upconverted emission (**Fig. 6a**). Singlet excitons photogenerated in donor (DIB-SQ with $S_1 = 1.7\ eV^{[34]}$) or acceptor (PCBM with $S_1 = 1.72$ eV[61], step 1 in **Fig. 6a**) dissociate by charge transfer across the D-A interface, forming either spin-singlet charge transfer (¹CT) states or immediate long-range charge separation (CS)[5-6, 62] (step 2). The non-geminate recombination of the initially separated charges leads to the formation of almost degenerate S = 0, ¹CT, and S = 1, ³CT states in a 1:3 ratio, as dictated by spin statistics (step 3). In materials with a low driving force for charge separation, geminate ³CT states can also be formed on the nanosecond timescale through the hyperfine interaction induced spin mixing ISC[62-65] (step 4). Due to the relatively long lifetime of the ³CT, (since their decay to the ground state is spin-forbidden), rubrene triplet states, which lie in close proximity to the BHJ, can be effectively populated. This occurs either via a direct triplet energy transfer from the ³CT states formed at the D-A interface (here DIB-SQ:PCBM, step 5) or indirectly through molecular triplets of the donor or acceptor molecules (here via molecular triplets in PCBM with $T_1$ energy of 1.5 eV[61]) if their energy is lower than that of the ³CT energy but greater than rubrene $T_1$ (1.14 eV[66], steps 6a,b). When a sufficient concentration of rubrene triplets is present, the annihilation of two triplets (TTA, step 7) can regenerate a rubrene singlet, leading to the observed upconverted emission (step 8). Although, the stepwise pathway via PCBM triplets (steps $6_{a,b}$) may incur additional losses, it may still contribute to an increase in the population of rubrene triplets due to the favorable alignment of energy levels. Specifically, the $T_1$ energy level of PCBM is approximately 0.36 eV higher than that of rubrene.

Further evidence of triplet-triplet annihilation upconversion mechanism is demonstrated by upconverted photoluminescence measurements in an external magnetic field (**Fig. 6b**). In upconversion (DIB-SQ:PCBM (1:3)/rubrene:DBP device excited at 690 nm), photoluminescence arises from the triplet–triplet annihilation (TTA) of rubrene triplets, which are populated via triplet energy transfer (TET) from the sensitizer. The observed magnetic field effect (MPL signal, defined as the relative change in photoluminescence intensity $MPL = \frac{PL(B)-PL(B=0)}{PL(B=0)}$) is negative and indicates an almost 10% decrease in upconverted emission at 300 mT field. This can be attributed to the reduction in the number of emissive triplet–triplet pairs with a singlet character in the encounter complex upon application of an external magnetic field, as described by Merrifield[63, 67]. Similar dependence of the MPL signal was observed in other solid-state upconversion systems with DBP doped rubrene annihilator[17]. On the other hand, when rubrene singlets are populated directly via optical excitation (rubrene:DBP sample excited at 450 nm), the external magnetic field affects the singlet fission rate[68]. Accordingly, photoluminescence increases by more than 4% on the application of 300 mT field.

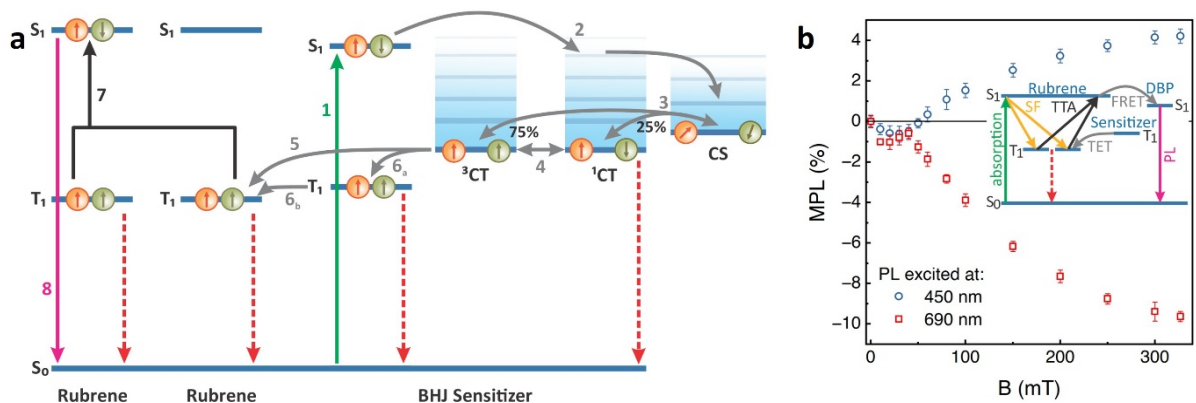

**Figure 6. Photon upconversion mechanism.** (a) State diagram illustrating photophysical processes involved in the population of rubrene triplet states followed by TTA-induced photon upconversion. $S_1$ and $T_1$ represent the lowest-energy molecular singlet and triplet excitons, respectively. CT denotes the energy of the relaxed, Coulombically bound electron-hole pair across the heterojunction and CS denotes the charge separated state. The conversions between excited state species are shown by solid grey lines, while recombination channels are indicated by red dashed lines. A photon absorption event (step 1) in the BHJ sensitizer is followed by singlet exciton dissociation and the formation of $^1$CT states that separate into free charges (CS) (step 2). Non-geminate recombination of electrons and holes leads to the formation of $^1$CT and $^3$CT states in a 1:3 ratio (step 3). Intersystem crossing (ISC) in CT states due to hyperfine interactions can occur on a nanosecond timescale (step 4). Triplet energy transfer, either directly from long-lived $^3$CT states (step 5) or indirectly through molecular triplets in either D or A (steps $6_{a,b}$), populates the triplet states in rubrene. Finally, two triplets can recombine to generate a singlet via TTA (step 7), which leads to the energy upconverted emission (step 8). (b) Magnetic field effect on photoluminescence (MPL signal) for rubrene:DBP film excited at 450 nm (blue dots, classical photoluminescence) and for DIB-SQ:PCBM (1:3)/rubrene:DBP device excited at 690 nm (red squares, upconverted photoluminescence), respectively. Error bars represent two standard deviations of at least ten data points. The inset shows a simplified state diagram illustrating the photophysical processes leading to photoluminescence in the rubrene:DBP system. The change in the number of triplet-triplet pair states with partial spin-singlet character arises from magnetic field-induced changes in the rate constant for triplet-triplet annihilation or singlet fission.

## 3. Conclusions

In summary, we have demonstrated a promising approach to harvesting low-energy photons for sensitized triplet generation using a bulk-heterojunction composition that permits fine tuning of photoexcited states by varying the amount of fullerene, enabling the effective buildup of rubrene triplet states to levels sufficient for TTA-induced photon upconversion. This approach leverages the prevalent back electron transfer in donor-acceptor BHJ systems, typically considered as a terminal loss channel in photovoltaics, and repurposes it into a triplet sensitizer to drive beneficial TTA-based upconversion in rubrene. Time-resolved measurements revealed competing coupling pathways among CS, $^3$CT, and molecular triplets, allowing the NIR excitation energy to ultimately funnel through the $T_1$ state of the annihilator, thereby facilitating UC in the solid-state.

The proposed upconversion system benefits from the triplet energy transfer occurring from sensitizer $^3$CT states directly and indirectly via molecular triples, instead of an intermediate step of CT states formed at sensitizer/annihilator interface, as shown in previous studies. This mechanism unlocks the potential of a wide range of low-bandgap materials, which have previously been treated unsuitable due to inefficient photogeneration as NIR sensitizers, for triplet state formation in upconversion energy cascades. The relatively high upconversion quantum yield of 1.36% at modest light intensities bears great promise for the technological implementation of photon upconversion in field deployable optoelectronic devices, paving the way for their integration into commercial photon imagers.

## 4. Experimental Section

### *UC device fabrication*

Glass substrates were sequentially cleaned in an ultrasonic bath with Alconox detergent (Sigma-Aldrich), deionized (DI) water, acetone, ethanol, and isopropanol (15 min each), dried using a nitrogen gas gun, and shortly annealed in air to remove any trace of solvent. Subsequently, substrates were treated in UV-ozone for 10 min and transferred into a nitrogen-filled glovebox. DIB-SQ (Sigma-Aldrich) was mixed with PCBM (Nano-C) at 1:0, 1:0.5, 1:1, 1:2, 1:3, 1:4, 1:5, and 0:3 weight ratios with overall concentrations of 5, 7.5, 10, 15, 20, 25, 30, and 15 mg/mL in anhydrous chlorobenzene (99.8%, Sigma-Aldrich). The solutions were stirred overnigh at 60°C in sealed vials. The active layers were spin-coated at 1500 rpm for 60 s. The samples were then annealed at 120°C for 10 min and placed in an evaporation chamber. A 30 nm thick layer of rubrene (sublimed grade, 99.99%, Sigma-Aldrich) or rubrene doped with DBP (98%, Sigma-Aldrich) at 0.5 vol. %, was grown under high vacuum (~$10^{-6}$ mbar) at a rate of 0.2-0.5 Å/s. DBP doping was achieved via co-evaporation by adjusting the deposition rates accordingly. Finally, the samples were sealed inside a glovebox using a 2-part epoxy resin (Araldite, Selleys) and a glass slide to protect the active layers against oxygen and moisture. For the TA and PLQY measurements, 1 mm fused silica slides were used instead of glass.

### *OPVs fabrication and characterization*

The ITO-coated glass substrates were cleaned in a similar manner to the glass substrates described above. PEDOT:PSS (AI 4083, Ossila) was spin-coated at 5000 rpm for 60 s on the ITO substrates and annealed at 110°C for 15 min. The active layers were spin-coated from DIB-SQ (5 mg/mL), PCBM (10 mg/mL), DIB-SQ:PCBM (1:3, 20 mg/mL) solutions in anhydrous chlorobenzene at 1500 rpm for 60 s. The samples were then annealed at 120°C for 10 min and placed in an evaporation chamber. The devices were completed by thermally evaporating a cathode consisting of 1 nm LiF and 100 nm Al through a shadow mask at a rate of 0.2-0.5 Å/s. The external quantum efficiency (EQE) was measured using a QUANTX-300 quantum efficiency measurement system (Newport) with 350-1800 nm range, at 0 V bias.

### *Steady-state absorption and photoluminescence*

Absorption measurements were performed using a Cary 60 UV-Vis and Cary 5000 UV-vis-NIR spectrophotometers (Agilent) in transmission mode. For the encapsulated samples, the spectrum of a dummy device with no active layers was used for background subtraction. Photoluminescence spectra were recorded using a Cary Eclipse Fluorescence Spectrometer (Agilent). The same instrument was used for excitation wavelength-dependent upconversion measurements. The obtained emission spectra were integrated from 520 to 600 nm. Long-pass filters with a cut-on wavelength of 590 and 630 nm (OG590 and RG630, SCHOTT) were used on the excitation side to remove any higher order harmonics.

*Time-Resolved Photoluminescence*

TRPL dynamics were collected in front-face detection geometry using a time-correlated single-photon counting (TCSPC, DeltaFlex, HORIBA) system operated in multi-channel scaling (MCS) mode. A single-photon detection module (PDD-920) collected the signal, which was filtered by a time-domain monochromator (TDM-1200). A nanosecond-pulsed laser diode, (NPL64C, Thorlabs), $\lambda = 640\ nm$, operated at a 50 kHz repetition rate, with ~20 ns pulse width and pulse energy of 126 nJ was used as the excitation source. The instrument response function (IRF) had a FWHM of ≈85 ns. A short-pass edge filter cutting off at $\lambda = 600\ nm$ (600FL07-50S, Andover) was used to reject any scattered pump light, and a long-pass edge filter with a cut-on wavelength of $\lambda = 600\ nm$ (600FH90-50S, Andover) was used on the excitation side. Mono-exponential growth and decay functions were fitted to extract upconverted photoluminescence lifetimes from the resulting decay curves using commercially available software (EzTime, HORIBA): $I_{UC}(t) \propto e^{-t/\tau_d} - e^{-t/\tau_g}$.

*Transient absorption spectroscopy*

Femtosecond transient absorption (fs-TA) measurements were undertaken using an amplified laser system (Spitfire ACE, Spectra-Physics) as the excitation source, based on an 800 nm Ti:Sapphire seed (MaiTai, Spectra-Physics) and 532 nm DPSS Nd:YLF pump laser (Ascend 60, Spectra-Physics), delivering ca. 150 fs 800 nm laser pulses at a 1 kHz repetition rate. Approximately 0.2 mJ of this output was attenuated and focused onto a 2 mm sapphire plate to generate a white light continuum probe pulse in the vis region ≈450-820 nm and NIR region ≈850-1500 nm. The remainder of the laser fundamental was coupled to an OPA system (Topas Prime, Light Conversion) delivering femtosecond tunable excitation pulses at 688 nm (0.15-0.9 μJ pulse$^{-1}$) and the pump pulse polarization was set to a magic angle with respect to the probe using a 10 mm Glan–Taylor polarizer (GT10, Thorlabs). Transient difference spectra at variable time delays were collected using a CCD based spectrometer (Helios, Ultrafast Systems). The IRF had a FWHM of ≈200 fs, as measured experimentally by a Gaussian fit to the scattered laser excitation profile, and all spectra were corrected for the chirp of the probe pulses. For nanosecond-transient absorption measurements (ns-TA), a white light continuum from ≈410 to 1550 nm was generated by using a pulsed Nd:YAG based Leukos-STM super continuum light source, the timing of which was controlled electronically using the sync out of the amplified laser system. The IRF for this setup dictated by the electronic timing resolution was $\approx$100 ps. Ground and excited state difference spectra at various delay times were measured using a sub nanosecond, CCD based spectrometer (EOS, Ultrafast Systems).

The resulting data was analyzed using commercially available software (Igor, Version 6.1.2.1, Wavemetrics, Inc). Kinetic traces averaged every 10 nm were extracted from the raw data and fit to a fully parallel global model. The singular value decomposition (SVD) was used to determine the number of exponential functions used in each model. Data from visible and NIR probe experiments was fit separately and used to build the overall kinetic model. The excitation pulse energy was kept at 0.15 μJ for neat DIB-SQ, and between 0.5-0.9 μJ for composite films.

*Upconversion threshold*

The dependence of the upconversion excitation intensity was measured using the same setup used for TRPL. A 690 nm continuous-wave (cw) laser diode (HL6738MG, Thorlabs) was used as an excitation source. A short-pass edge filter cutting off at $\lambda = 650\ nm$ (650FL07-50S, Andover) was used to reject any scattered pump light, and a long-pass edge filter with a cut-on wavelength of $\lambda = 650\ nm$ (650FH90-50S, Andover) was used on the excitation side. The obtained emission spectra were integrated from 550 to 620 nm.

*PLQY measurements*

The PLQY of rubrene:DBP film was measured by a spectralon lined integrating sphere of 5.3" interior diameter (Labsphere) with the output fiber coupled to a Flame spectrometer (OceanOptics), which was calibrated using a tungsten halogen vis-NIR radiometrically calibrated light source with a wavelength range from 350 nm to 2.4 μm (OceanOptics). Samples were excited by a 450 nm cw laser diode (L450P1600MM, Thorlabs). The De Mello method[50] was used for data analysis.

The upconversion quantum yield ($\Phi_{UC}$) was measured using a relative method, and was calculated by the following equation (1):

$$\Phi_{UC} = \left(\frac{I_{UC}}{I_{std}}\right)\left(\frac{P_{std}}{P_{UC}}\right)\left(\frac{1-10^{-A_{std}}}{1-10^{-A_{UC}}}\right)PLQY_{std}, \tag{1}$$

where $A$ is the absorbance at the excitation wavelength. $I$ is the PL intensity and $P$ is the number of photons irradiated from the excitation light source. Subscripts UC and std refer to the UC emission and the standard sample, respectively. A rubrene:DBP film with PLQY of 47.5±2.4%, excited at 450 nm by a cw laser diode (L450P1600MM, Thorlabs) was used as a standard. Measurements were carried out on the same setup as previously described in the excitation intensity experiments.

*Magnetic field effect measurements*

Samples were excited with either a 450 nm cw laser diode (L450P1600MM, Thorlabs) for photoluminescence measurements, or 690 nm cw laser diode (HL6738MG, Thorlabs) for upconversion measurements, and the emission was measured using a Flame spectrometer (OceanOptics). A long-pass edge filter cutting off at $\lambda = 500\ nm$ (500FH90-50S, Andover), or a short-pass edge filter cutting off at $\lambda = 650\ nm$ (650FL07-50S, Andover) was placed in front of the detector to block laser scatter during PL or UC measurements, respectively. The magnetic field was applied parallel to the device plane using an electromagnet (DXWD-50, DexinMag). The magnetic field strength at the sample position was measured with a Gaussmeter (DX-150, DexinMag). The magnetic field effect on photoluminescence, denoted as MPL, is defined as a relative change of photoluminescence intensity being the function of the field of strength B applied to the sample $MPL = \frac{PL(B) - PL(B=0)}{PL(B=0)}$.

**Acknowledgments**

We would like to acknowledge Defence Science and Technology Group for funding this research and Central Analytical Research Facility (CARF) at QUT for the use of laboratories and experimental facilities. M.K. acknowledges support from the Centre for Materials Science, QUT under "QCMS Boost" project. E.G.M. and A.K.P. acknowledge Australian Research Council grant (LE210100124) that supported the ultrafast spectroscopy facility at UQ.

**Author contributions**

M.K., A.K.P. and D.D. conceived the idea. M.K. developed the upconversion platform, planned and performed all characterization. A.R.I. and M.K. performed the TA measurements under the supervision of E.G.M. M.K. and A.K.P. wrote the first draft of the manuscript. All the authors analyzed the results and contributed to finalizing the manuscript draft. A.K.P. and D.D. supervised the work.

# Charge Transfer States in Donor-Acceptor Bulk-Heterojunctions as Triplet-Triplet Annihilation Sensitizer for Solid-State Photon Upconversion

Maciej Klein,[1,2*] Alexander R. Ireland,[3] Evan G. Moore,[3*]

Dennis Delic[4] and Ajay K. Pandey[1,2*]

[1] *School of Electrical Engineering and Robotics, Queensland University of Technology, Brisbane, QLD 4001, Australia*

[2] *Centre for Materials Science, Queensland University of Technology, Brisbane, QLD 4001, Australia*

[3] *School of Chemistry and Molecular Biosciences, The University of Queensland, Brisbane, QLD 4072, Australia*

[4] *Defence Science and Technology Group, Edinburgh, SA 5111, Australia*

*Correspondence to: maciej.klein@qut.edu.au, egmoore@uq.edu.au, a2.pandey@qut.edu.au*

**Brief background of TTA derived UC**

In TTA-based UC, low-energy input photons are stored as triplet excitons (bound e-h pairs), the form of long-lived molecular states in the annihilator. These states are essentially 'dark states' due to the forbidden nature of the direct optical transition between the S = 0 ground state and the S = 1 triplet levels[1]. Therefore, TTA-based upconversion usually require pairing up of the photoemitter with a suitable sensitizer to ensure efficient absorption of NIR photons[2], leading to formation of a photoexcited singlet state (S₁) followed by relaxation into a triplet state (T₁) via intersystem crossing (ISC)[3] (induced by strong spin-orbit coupling due to the heavy atom effect)[4-5]. This is permissible via a Dexter type triplet-triplet energy transfer (TET) between the sensitizer and annihilator[6]. The fusion of triplets via TTA leads to formation of a high-energy singlet state in the annihilator molecule, which can now radiatively decay to the ground state emitting a high-energy photon, conserving roughly isoergic conditions, $E(S_1) = 2E(T_1)$.

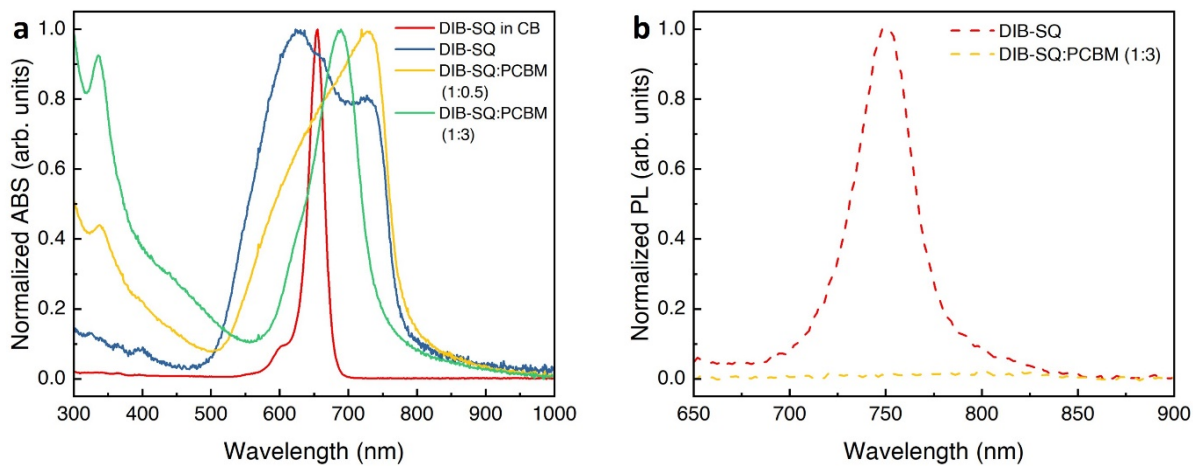

**Figure S1.** Normalized (a) absorption (ABS) and (b) photoluminescence (PL) spectra of DIB-SQ and DIB-SQ:PCBM bulk heterojunction films on a quartz substrate, and DIB-SQ in chlorobenzene solution (DIB-SQ in CB). The excitation wavelength was 580 nm.

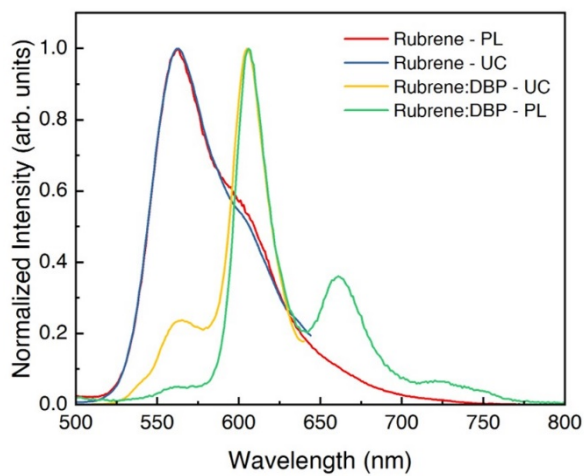

**Figure S2.** Normalized photoluminescence (PL) spectra of pristine rubrene and doped rubrene:DBP films excited at 450 nm, along with upconversion (UC) spectra of DIB-SQ:PCBM (1:3)/rubrene and DIB-SQ:PCBM(1:3) /rubrene:DBP excited at 690 nm.

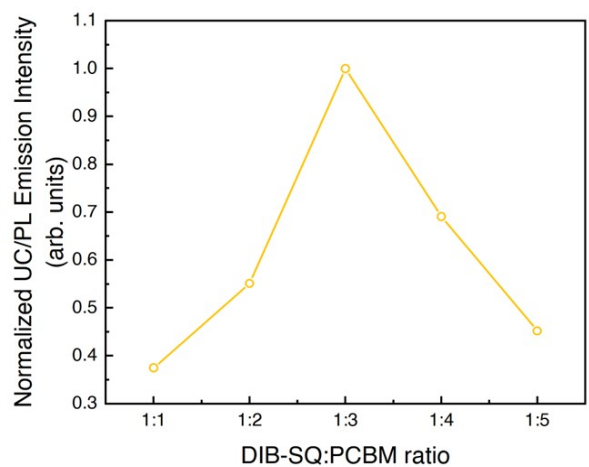

**Figure S3.** Effectiveness of the upconversion defined as a ratio of the upconverted photoluminescence intensity (excited at maximum sensitizer absorption) to direct rubrene photoluminescence (excited at 465 nm), UC/PL, for varying DIB-SQ to PCBM weight ratio.

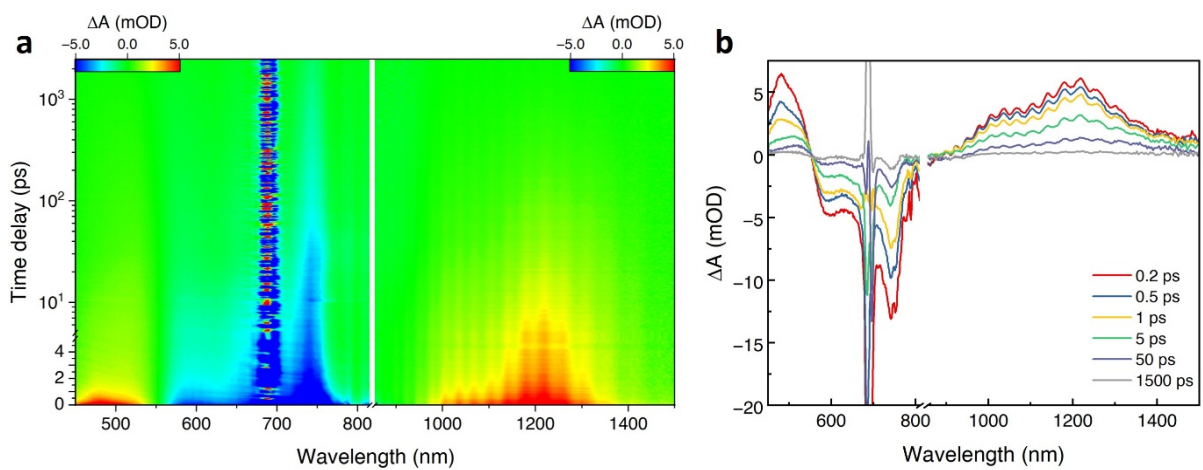

**Figure S4.** (a) 2D femtosecond-transient absorption contour map and (b) related TA spectra of DIB-SQ:PCBM (1:0.5) film at different probe delay times. Small spectra oscillations, visible mainly in the NIR range are caused by an optical cavity effect formed by the close proximity of substrate and encapsulation slides. Some scattered light from the excitation pulse is observed at $688 \pm 10$ nm.

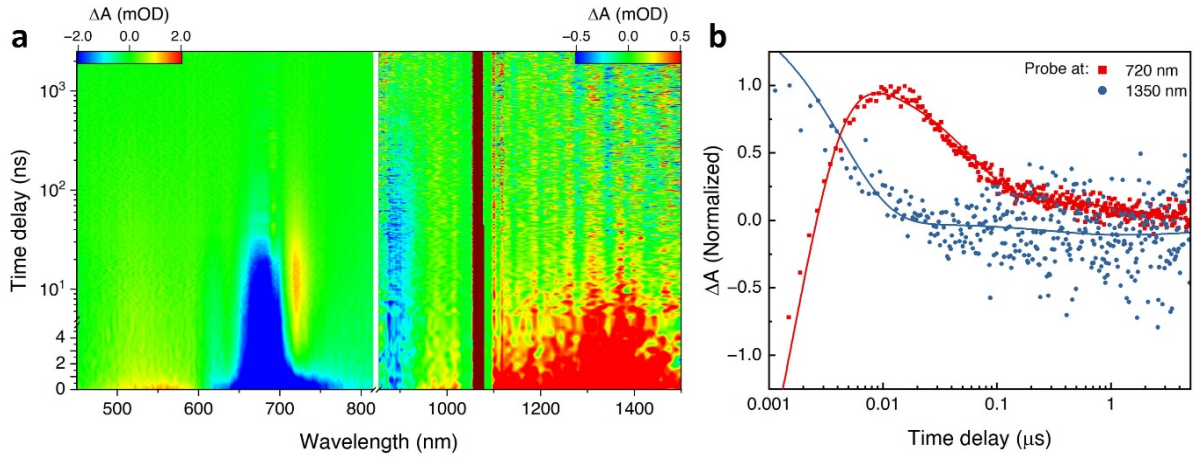

**Figure S5.** (a) 2D nanosecond-transient absorption contour map of DIB-SQ:PCBM (1:3) film. (b) Normalized TA decay kinetics of DIB-SQ:PCBM (1:3) film monitored at 720 and 1350 nm. The solid lines represent the associated kinetic fitting curves extracted from global analysis.

## Estimation of CT state energy

The CT state energy ($E_{CT}$) was determined by fitting the low-energy absorption tail of the reduced EQE spectrum and the high-energy tail of the reduced emission spectrum. Both fits were performed using a Gaussian function with amplitude $A_0$, centered around $E_c$ and given by the equation: $A(E) = A_0 \exp\left[-(E - E_c)^2/(2\sigma^2)\right]$, where σ is the standard deviation (Ref. 56 in the main manuscript). The $E_{CT}$ is provided by the crossing energy of the two fitted spectra, after normalization to the maximum of the corresponding peaks. The reduced EQE and emission spectra were obtained by dividing the EQE(E) spectrum (of DIB-SQ:PCBM(1:3) shown in Fig. S6a) by the energy, E, and the emission spectrum (extracted from the nanosecond-transient absorption spectra of DIB-SQ:PCBM at 2 ns probe delay time, shown in Fig. S5) by $E^3$ [7]. The intersection between the fitted reduced PL and EQE spectra yields the $E_{CT}$ energy of 1.44 eV (~861 nm). The peak energy of the fitted EQE and PL curves are 1.46 and 1.36 eV, respectively.

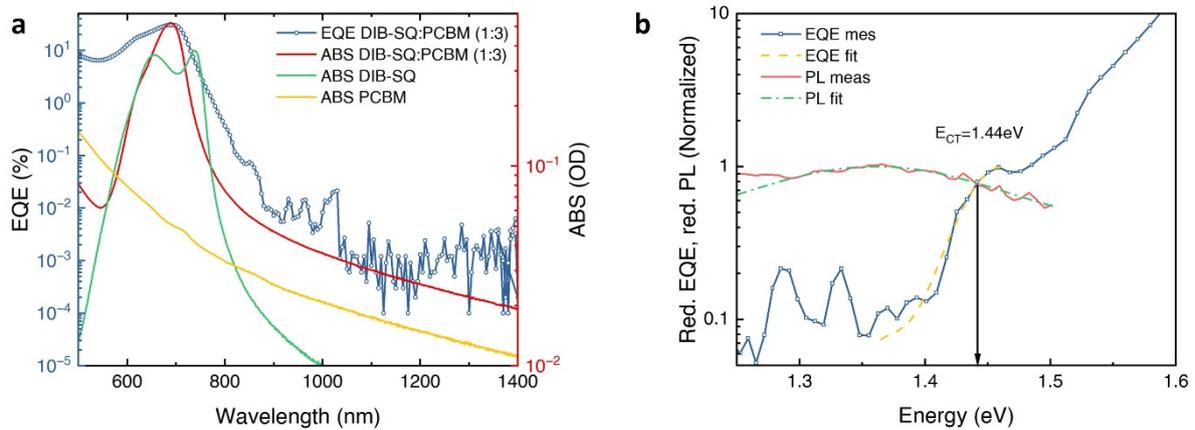

**Figure S6.** (a) External quantum efficiency (EQE) of a OPV with DIB-SQ:PCBM (1:3) active layer overlaid with absorption (ABS) spectra of neat DIB-SQ, neat PCBM, and bulk-

heterojunction DIB-SQ:PCBM (1:3) films on a quartz substrate. <mark>(b) Reduced (red.) and normalized high-sensitivity EQE of the same OPV device, alongside reduced and normalized PL spectra extracted from ns-TA. Both spectra were fitted with the Gaussian function.</mark>

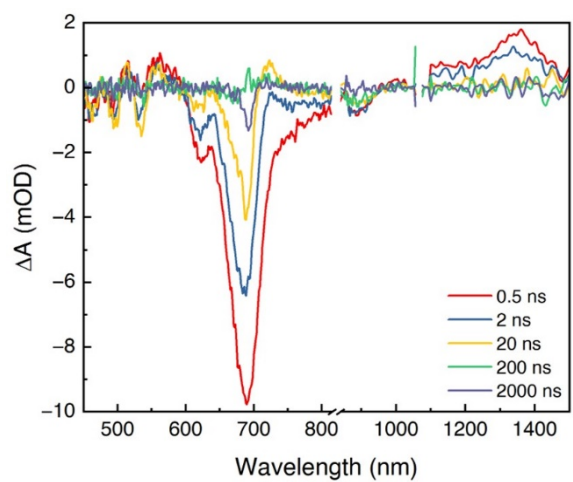

**Figure S7.** Nanosecond-transient absorption spectra of DIB-SQ:PCBM (1:3)/rubrene system at different probe delay times. Some scattered light from the excitation pulse is observed at 688 ± 10 nm.

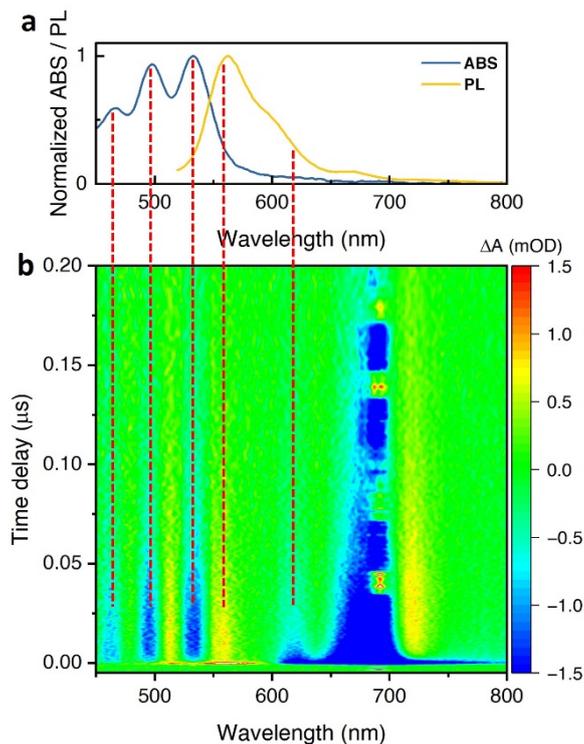

**Figure S8.** (a) Normalized absorption (ABS) and photoluminescence (PL) spectra of thermally evaporated rubrene film on quartz substrate. (b) 2D nanosecond-TA contour map of DIB-SQ:PCBM (1:3)/rubrene upconversion system deposited on quartz substrate. The dashed lines are guides to the eye, indicating the positions of the respective absorption and emission bands.